\documentclass[aps,prl,twocolumn,showpacs,superscriptaddress]{revtex4}

\usepackage{graphicx}

\usepackage{amsmath}

\begin{document}
 
\title{First-Principles Study of Integer Quantum Hall Transitions in
Mesoscopic Samples}

\author{Chenggang Zhou}

\affiliation{ Department of Electrical Engineering, Princeton
  University, Princeton, New Jersey 08544, USA}

\author{Mona Berciu}

\affiliation{Department of Physics and Astronomy, University of
  British Columbia, Vancouver, BC V6T 1Z1, Canada } \date{\today}
 
\begin{abstract}

We perform first principles numerical simulations to investigate
resistance fluctuations in mesoscopic samples, near the transition
between consecutive Quantum Hall plateaus. We use six-terminal
geometry and sample sizes similar to those of real devices. The Hall
and longitudinal resistances extracted from the generalized Landauer
formula reproduce all the experimental features uncovered recently. We
then use a simple generalization of the Landauer-B\"uttiker model,
based on the interplay between tunneling and chiral currents -- the
co-existing mechanisms for transport -- to explain the three distinct
types of fluctuations observed, and identify the central region as the
critical region.

\end{abstract}


\maketitle

Although the Integer Quantum Hall Effect (IQHE) is a generally well
understood phenomenon, recent experiments on mesoscopic
samples~\cite{Peled1,Peled2} uncovered unexpected behavior in the
seemingly noisy fluctuations of the Hall ($R_H$) and longitudinal
($R_L$) resistances. Previously, fluctuations in resistance had been
observed in mesoscopic samples with a phase coherence length $L_\phi$
larger than the sample size~\cite{Timp, Simmons, Cobden}; they are
totally random, similar to universal conductance
fluctuations~\cite{Fisher}.  By contrast, Peled {\em et.~al.} find
\cite{Peled1,Peled2} that the transition between the $n^{th}$ and
$(n+1)^{st}$ plateaus of the IQHE has three distinct regimes: (i) on
the high-$B$ side, both $R_H$ and $R_L$ have large but correlated
fluctuations, such that $R_L+R_H = h/n e^2$; (ii) for intermediate $B$
values, $R_H$ and $R_L$ continue to exhibit fluctuations, but their
sum is no longer constant; and (iii) on the low-$B$ side, $R_H =
h/(n+1)e^2$ has no fluctuations, whereas $R_L$ still does. Moreover,
$R_L+R_H=R_{2t}$ holds for all $B$ values~\cite{Peled2}.  Changing the sign of the
magnetic field $B\rightarrow -B$ also has interesting consequences, as
discussed later.  For $n=0$, regions (i) and (ii) are replaced by the
transition to the insulating phase~\cite{Peled1}. In this Letter, we
explain the physics behind these observations in a unified theory, and
analyze the implications for further experimental and theoretical
study.

The relation $R_L+R_H=R_{2t}$ was first proposed by Streda {\em
et. al.}~\cite{MacDonald}, while the fluctuations of regime (iii) are
reminiscent of Jain and Kivelson's theory on the resistance
fluctuations of narrow samples~\cite{Jain}. These theories were
questioned by B\"uttiker~\cite{Buttiker1}, based on formulas
appropriate for a four-terminal geometry~\cite{Buttiker2}. We take an
approach similar to B\"uttiker's and use the multi-probe Landauer
formula~\cite{Buttiker2, Baranger} to calculate the resistances
measured experimentally. However, we mirror the experiments by
including six contacts in our model, namely the four voltage probes
plus the source and the drain for the electrical current. Our model
enables us to calculate both $R_L$ and $R_H$, and reveals the very
rich physics underlying the observation of the mesoscopic IQHE.

\begin{figure}[b]
\includegraphics[width=\columnwidth]{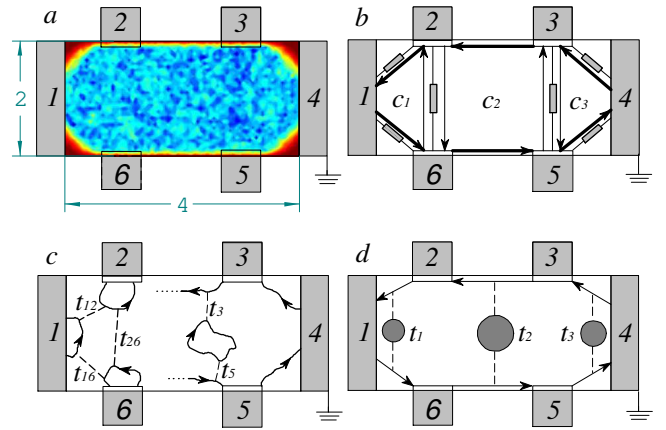}
\caption{ (color online) (a) Typical potential $V_d+V_b$ of a
$4\mu$m$\times 2\mu$m sample. (b) Chiral (arrows) and tunneling
(resistors) currents in our model. This direction of chiral currents
corresponds to $B$ entering the page. (c) Some semi-classical current
distributions parameterized by our model. (d) Jain-Kivelson tunneling
for high-$\nu$. See text for further details. }
\label{fig1}
\end{figure}

The response function of the system is a $6\times6$ conductance matrix
$\hat{g}$, with which the current-voltage relation reads $I_\alpha
=\sum_\beta g_{\alpha \beta} V_\beta$. Here, $I_\alpha$ is the
out-going current on the contact $\alpha=1, \cdots, 6$ and $V_\alpha$
is the corresponding voltage. For a mesoscopic Hall bar, $\hat{g}$
characterizes the electrical response (since the system is
inhomogeneous, local quantities such as the current density are not
conceptually well-defined).  $\hat{g}$ is calculated~\cite{Baranger}
by solving a multi-channel scattering problem: $g_{\alpha,\beta\ne
\alpha} = \frac{e^2}{h} \sum_{i,j}\left| t_{\alpha i, \beta
j}\right|^2$, where $t_{\alpha i, \beta j}$ is the transmission
amplitude from the $j^{th}$ transverse channel of contact $\beta$ into
the $i^{th}$ transverse channel of contact $\alpha$ for an electron at
the Fermi energy $E_F$.  Due to the absence of charge accumulation and
to gauge invariance, $\sum_\alpha g_{\alpha \beta} = \sum_\beta
g_{\alpha \beta} = 0$.  As a result, diagonal $g_{\alpha\alpha}$ are
uniquely determined by $g_{\alpha \alpha} =-\sum_{\beta\ne
\alpha}g_{\alpha \beta}=-\sum_{\beta\ne \alpha}g_{\beta \alpha
}$. This imposes a constraint on the off-diagonal elements of
$\hat{g}$ for each $\alpha$.

Our model is sketched in Fig.~\ref{fig1}(a). Six perfectly conducting
contacts are linked to a 4$\mu$m$\times2\mu$m sample which has a
disorder potential $V_d({\bf r})$ and a background potential $V_b({\bf
r})$. $V_d({\bf r})$ is a sum of random Gaussian scatterers generating
elastic scattering inside the sample, while $V_b({\bf r})$ confines
the electrons to the sample and creates edge states.  In our
simulations, we include $L_xL_yB/\phi_0 \sim 10^4$ states of the
lowest Landau level (LLL), where $L_xL_y$ is the area of the sample
and $\phi_0=h/e$ is the magnetic flux quantum. The sample Hamiltonian
is a large, sparse matrix $H_{nm} =
\left<\psi_n|V_b+V_d|\psi_m\right>$. Contacts are modeled by ensembles
of one-dimensional tight-binding chains attached to localized
eigenstates on the corresponding edges of the sample. We have verified
the multi-probe Landauer formula for our model using the linear
response theory. This derivation and further modeling details will be
reported elsewhere~\cite{Zhou}.  For a given magnetic field $B$, we
numerically solve the scattering problem for different values of the
Fermi energy and obtain $\hat{g}$. The filling factor $\nu$ is also a
function of $E_F$, and thus we can find the dependence of the
conductance matrix $\hat{g}$ on $\nu$.

The resistances are then computed from $\hat{g}$. In the usual setup
the current is injected in the source and extracted in the drain
$-I_1=I_4=I$; $\hat{I}_{14} = \begin{pmatrix} -I&0&0&I&0&0
\end{pmatrix}^T$. Without loss of
generality we set $I=1$ and $V_4 = 0$. The other five contact voltages
are uniquely determined from $\hat{I}_{14} = \hat{g}\cdot \hat{V}$.
We define two longitudinal resistances $R^L_{14,23} = (V_2 -V_3)/I =
V_2-V_3$, $ R^L_{14,65} = V_6-V_5$, and two Hall resistances
$R^H_{14,62} = V_6-V_2$, $ R^H_{14,53} = V_5 -V_3$.

\begin{figure}[t]
\includegraphics[width=\columnwidth]{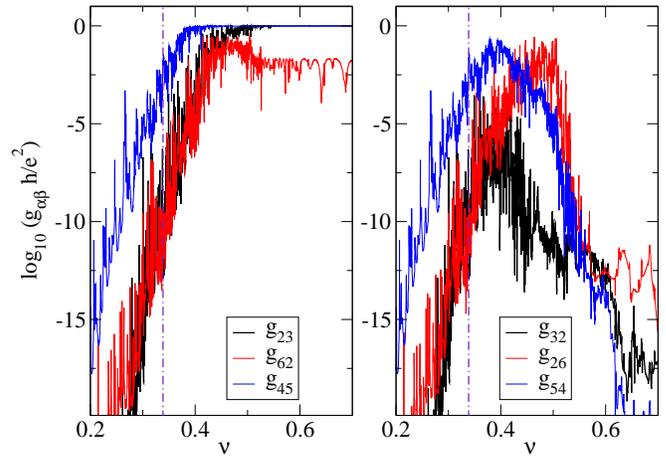}
\caption{(color online) Representative conductance matrix elements, in
  units of $e^2/h$, as a function of the filling factor $\nu$. The
  left (right) panel shows $g_{23}, g_{45}$ and $g_{62}$, respectively
  $g_{32}$, $g_{54}$ and $g_{26}$ characterizing transport in the
  (against the) direction of the edge currents. Results are almost
  identical on the left of the dot-dashed line, but different on its
  right. }
\label{fig2}
\end{figure}

In Fig.~\ref{fig2}, we plot representative matrix elements $g_{\alpha
\beta}$ as a function of $\nu$. For $\nu>0.5$, we find $g_{\alpha,
\alpha+1} \rightarrow e^2/h$ (if $\alpha=6$, $\alpha+1=1$), with all
other off-diagonal matrix elements vanishing. In other words, all
electrons leaving contact $\alpha+1$ arrive at contact $\alpha$.  It
follows that here
\begin{equation}
\label{g0}
g_{\alpha\beta} \rightarrow g^{(0)}_{\alpha \beta} = \frac{e^2}{h}
\left(-\delta_{\alpha \beta} + \delta_{\alpha+1,\beta} +\delta_{\alpha
6}\delta_{\beta 1}\right).
\end{equation}
>From $\hat{I}_{14} = \hat{g}^{(0)}\cdot \hat{V}$ we find $V_5 = V_6 =
h/e^2$, $V_2 = V_3 = 0$, thus $ R^H_{14,62} = R^H_{14,53} = h/e^2$,
$R^L_{14,23} = R^L_{14,65} = 0$.  This shows that the first quantized
plateau is due to the chiral edge currents [shown as oriented thick
lines in Fig.~\ref{fig1}(b)], which become established for $\nu
>0.5$. Variations of $\hat{g}(\nu)$ from $ \hat{g}^{(0)}$ give rise to
fluctuations in the resistances.  From Fig.~\ref{fig2} we also see
that if $\nu < \nu_c$ (indicated by the vertical line), $g_{\alpha
\beta}\approx g_{\beta \alpha}$ with high accuracy, i.~e. $\hat{g}$ is
symmetric. For $\nu>\nu_c$, $\hat{g}$ is no longer symmetric. The
reasons for this behavior and its consequences are discussed later.

\begin{figure}[t]
\includegraphics[width=\columnwidth]{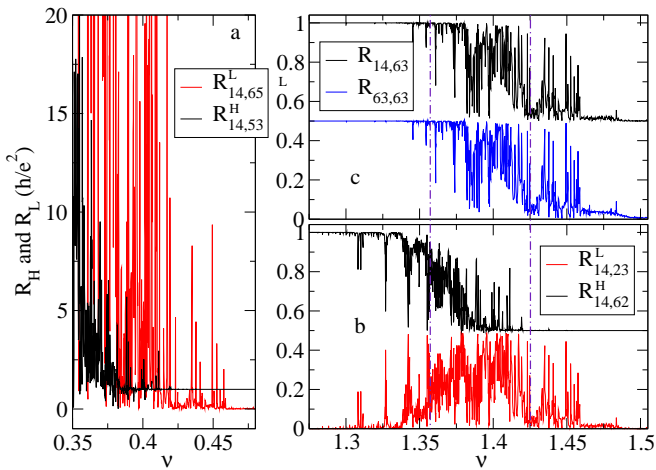}
\caption{(color online) $R_L$ and $R_H$ calculated from the
  conductance matrix displayed in Fig.~\ref{fig2}, in units of
  $h/e^2$.  (a) Transition from the insulator to the first IQHE
  plateau in the LLL. (b) Transition from the first to the second IQHE
  plateaus. (c) The sum $R_L+R_H$ of the resistances shown in (b), and
  $R_{2t}-0.5{h/e^2}$. Vertical lines indicate the boundaries of the
  critical region. See text for further details. }
\label{fig3}
\end{figure}

Using the conductance matrix $\hat{g}(\nu)$ plotted in
Fig.~\ref{fig2}, we now compute the values of the various resistances
as a function of $0<\nu<1$. Fig.~\ref{fig3}(a) shows a pair $R_L$ and
$R_H$. Three different regimes are clearly seen: for $\nu>0.46$,
$R_H=h/e^2$ and $R_L=0$, corresponding to the first IQHE plateau. For
$0.42<\nu<0.46$, $R_L$ exhibits large fluctuations, however $R_H$ is
still well quantized. This is precisely the type of behavior observed
in Ref.~\onlinecite{Peled1}. For $\nu < 0.42$ the transition to the
insulating phase occurs, and both resistances increase sharply. The
fluctuations are very large and narrow because the calculation is done
at $T=0$. At finite $T$, the peaks are smeared out.

The transition $1 < \nu < 2$ can also be simulated using the same
$\hat{g}(\nu)$ matrix of the LLL. Similar to Ref.~\onlinecite{Shahar},
we assume that the completely filled spin-up LLL contributes its
background chiral edge current. As a result, we simply add
$\hat{g}^{(0)}=\hat{g}(\nu= 1)$ of Eq.~(\ref{g0}) to the values of
$\hat{g}(\nu)$ describing the partially filled spin-down LLL. Although
the two LLLs have different spins, the contacts mix electrons with
both spins in equilibrium, justifying this addition.  Resistivities
$R^H_{14,62}$ and $R^L_{14,23}$ computed for
$\hat{g}^{(0)}+\hat{g}(\nu)$ are shown in panel (b) of
Fig. \ref{fig3}, whereas in panel (c) we plot their sum. The three
regimes found experimentally~\cite{Peled2,private} are clearly
observed. At low-$\nu$ (high-$B$), the fluctuations of $R_H$ and $R_L$
are correlated, $R_L+R_H=h/e^2$. At high-$\nu$ (low-$B$) $R_H=h/2e^2$
is quantized while $R_L$ still exhibits strong fluctuations. In the
intermediate regime, both $R_H$ and $R_L$ have strong, uncorrelated
fluctuations.  The other pair, $R^H_{14,53}$ and $R^L_{14,65}$, also
exhibits the three regimes, although their detailed fluctuations are
different from $R^H_{14,62}$ and $R^L_{14,23}$. From over 20 different
simulations we found that the low-$\nu$ regime where $R_L+R_H=h/e^2$
is a very robust feature, although it is maintained up to different
values of $\nu$ in different samples. The high-$\nu$ regime with
fluctuations in $R_L$ and quantized $R_H$ is seen frequently. However,
when strong direct tunneling occurs between the source or the drain
and their nearby voltage probes, $R_H$ also fluctuates. Such strong
tunneling is an artifact of our simulation~\cite{note1}. We suppress
it by isolating the source and drain from nearby contacts with
triangular potential barriers in the corners of the sample [see
Fig.~\ref{fig1}(a)]. Figure~\ref{fig3}(c) also compares
$R^L_{14,23}+R^H_{14,62} = R_{14,63}$ with $R_{2t}=R_{63,63}$. [In the
setup for measuring $R_{2t}$, the current is
$\hat{I}_{63}=\begin{pmatrix} 0&0&1&0&0&-1
\end{pmatrix}^T$, and $R_{2t} = V_6-V_3$]. As found
experimentally~\cite{Peled2}, the two curves are very similar.

So far, we have demonstrated that our numerical simulations recapture
faithfully the experimental results. We now explain the underlying
physics using a simple but very general model. For the given
constraints and using logical induction, we find~\cite{Zhou} that
$\hat{g}$ can be decomposed as a sum over {\em loops} connecting
contacts: $\hat{g} = \sum
c(k_1,\cdots,k_n)\hat{r}(k_1,\cdots,k_n)$. Here, $c$ are positive
numbers and $\hat{r}(k_1,\cdots,k_n) =
\hat{l}(k_1,k_2)+\dots+\hat{l}(k_n,k_1)$, where $l_{\alpha\beta}(a,b)
= \frac{e^2}{h} \left(\delta_{\alpha a} \delta_{\beta b} -\frac{1}{2}
\delta_{\alpha a} \delta_{\beta a}-\frac{1}{2}\delta_{\alpha
b}\delta_{\beta b}\right)$ contributes to a single off-diagonal
element $g_{ab}$. A two-vertex loop $\hat{r}(a,b)$ describes a $h/e^2$
resistor between contacts $a$ and $b$.  Since
${r}_{\alpha\beta}(a,b)={r}_{\beta\alpha}(a,b)$, these terms are the
symmetric part to $\hat{g}$. The asymmetric part of $\hat{g}$
describes chiral currents, whose direction of flow is dictated by the
sign of $B$.  In particular, $\hat{r}(1,2,3,4,5,6)=\hat{g}^{(0)}$ of
Eq. \ref{g0} describes the edge currents of a LL, but shorter chiral
circuits may also develop at intermediate fillings $\nu$.

At low-$\nu$, all states are localized and transport in the LL can
only occur through tunneling. Consider the semi-classical case
sketched on the left side of Fig.~\ref{fig1}(c). Electrons can go from
2 to 1 either through direct tunneling (probability $t_{12}$), or they
can tunnel to the localized state near contact 6 and from there back
into 1, with probability $(1-t_{12})t_{26}t_{16}(1-t_{12})$. Electrons
can make any number of loops before entering 1, the total sum being
$p_{2\rightarrow 1}= \frac{h}{e^2}g_{12} =
[t_{12}+t_{26}t_{16}-2t_{12}t_{26}t_{16}]/(1-t_{12}t_{16}t_{26})$. Similar
arguments give $p_{1\rightarrow 2}=
\frac{h}{e^2}g_{21}=[t_{12}-t_{26}t_{16}-t_{12}t_{16}]/(1-t_{12}t_{16}t_{26})$.
$g_{16},g_{61},g_{26}$ and $g_{62}$ are computed similarly. We define
$r_{ab} = \min( p_{a \rightarrow b}, p_{b\rightarrow a}) >0$, and
$c_{ab}= \max( p_{a \rightarrow b}, p_{b\rightarrow a})-r_{ab} >0$. We
find $c_{12}=c_{26}=c_{61}=c= t_{12}t_{16}+t_{12}t_{26}+t_{16}t_{26}
+O(t^3)$; and up to $O(t^2)$, $r_{12}\approx t_{12} $, $r_{16}\approx
t_{16}$ and $r_{26}\approx t_{26}$. These processes contribute a total
of $r_{12}\hat{r}(1,2) +r_{16}\hat{r}(1,6)+r_{26}\hat{r}(2,6) + c
\hat{r}(1,2,6)$ to $\hat{g}$. The symmetric resistance terms, of order
$t$, are due to direct tunneling between contacts, and at low-$\nu$
they dominate the small chiral current, of order $t^2$. This explains
why for $\nu < \nu_c$, $\hat{g}$ is symmetric with small off-diagonal
components (see Fig.~\ref{fig2}).  At high enough $\nu$, edge states
connecting consecutive contacts appear. As already discussed, as $\nu
\rightarrow 1$, $\hat{g} \rightarrow
\hat{g}^{(0)}=\hat{r}(1,2,3,4,5,6)$. For intermediate $\nu$, shorter
chiral loops containing edge states can be established through
tunneling, as sketched on the right side of Fig.~\ref{fig2}(c). Assume
that an electron leaving contact 3 can tunnel with probabilities $t_3$
and $t_5$ to and out of a localized state, to join the opposite edge
current and enter 5. It follows that $p_{3\rightarrow 5} =
\frac{h}{e^2} g_{53} = t_3t_5/[1-(1-t_3)(1-t_5)]$, whereas
$p_{5\rightarrow 3} =0$ (no electron leaving 5 enters 3). Then
$r_{35}=0$ and the contribution to $\hat{g}$ is just $p_{3\rightarrow
5} \hat{l}(5,3)$. This term combines with parts of $\hat{l}(3,4)$ and
$\hat{l}(4,5)$ to create a chiral current $c \hat{r}(3,4,5)$, where
$c=p_{3\rightarrow 5}$. Physically, this represents the backscattered
current of the Jain-Kivelson model~\cite{Jain}.

In general, one has to sum over many types of competing processes,
involving both tunneling and edge states, but $\hat{g}$ can always be
decomposed into symmetric resistances plus chiral loops. Consider the
general form $\hat{g} = n \hat{g}^{(0)}+
r_{12}\hat{r}(1,2)+r_{16}\hat{r}(1,6)+r_{26}\hat{r}(2,6)+
r_{34}\hat{r}(3,4)+r_{45}\hat{r}(4,5)+r_{35}\hat{r}(3,5)+
c_0\hat{g}^{(0)}+c_1\hat{r}(1,2,6)+c_2\hat{r}(2,3,5,6)+c_3\hat{r}(3,4,5)
+c_4\hat{r}(1,2,3,5,6)+ c_5\hat{r}(2,3,4,5,6)$. The first term
describes the contribution of the $n$ completely filled lower LLs. All
other terms describe transport in the LL hosting $E_F$ [see
Fig.~\ref{fig1}(b)], with the restriction that there is no tunneling
between the left and right sides of the sample. This is justified
physically because tunneling between contacts far apart is vanishingly
small. The largest such terms, $r_{23}$ and $r_{56}$, are found to be
less than $10^{-4}$ [see e.g. Fig.~\ref{fig2} , where
$r_{23}=h/e^2\cdot\min(g_{32},g_{32})$]. We solve both
$\hat{I}_{14}=\hat{g}\cdot \hat{V}$ and $\hat{I}_{63}=\hat{g}\cdot
\hat{V'}$ and find the identity
$$ R_{14,63}=R_{63,63} = \frac{h}{e^2} \frac{1}{n + c_0 + c_2 + c_4 +
c_5}
$$ Since $R_{63,63}=R_{2t}$, whereas $R_{14,63} =
R^H_{14,62}+R^L_{14,23} =R^L_{14,65}+ R^H_{14,53}$, this means that $
R_{2t} = R_H + R_L $ irrespective of the value of the 12
parameters. In other words, this identity is obeyed for all $\nu$, in
agreement with Fig.~\ref{fig3}(c). (Adding $r_{23}$ and $r_{56}$ terms
leads to perturbationally small corrections~\cite{Zhou}). Here $n +
c_0 + c_2 + c_4 + c_5$ is the total chiral current flowing along the
$6\rightarrow 5$ and $3\rightarrow 2$ edges.  In particular, at
low-$\nu$ chiral currents in the LL hosting $E_F$ are vanishingly
small $c_0=c_2=c_4=c_5=0$ (there are no edge states established yet,
and pure tunneling contributions are of order $t^2$, as already
discussed. Below $\nu_c$, all $t < 10^{-4}$, see Fig.~\ref{fig2}). It
follows that here $R_L + R_H = h/ne^2$, explaining the perfect
correlations in the pattern fluctuations at low-$\nu$ of the two
resistances, observed both experimentally and numerically.

The high-$\nu$ regime with quantized $R_H$ and fluctuating $R_L$ can
also be understood easily. As already discussed, transport in the LL
hosting $E_F$ is dominated here by the edge states; tunneling between
opposite edge states (facilitated by localized states inside the
sample) creates backscattered currents, as in the Jain-Kivelson
model~\cite{Jain}. We sketch this situation in Fig.~\ref{fig1}(d),
where $t_1$, $t_2$ respectively $t_3$ include all possible tunneling
processes leading to backscattering on the corresponding pairs of edge
states. Reading the various probabilities off Fig.~\ref{fig1}(d), we
find that $\hat{g}= n\hat{g}^{(0)} + (1-t_1-t_2-t_3)\hat{g}^{(0)} +
t_2[\hat{r}(1,2,6)+\hat{r}(3,4,5)] + t_3 \hat{r}(1,2,3,5,6) + t_1
\hat{r}(2,3,4,5,6)$. The first term represents the contribution of the
lower $n$ completely filled LLs, the others are the forward and the
backscattered chiral currents in the LL hosting
$E_F$. $\hat{I}_{14}=\hat{g}\cdot \hat{V}$ is trivial to solve. We
find $R^H_{14,62} = R^H_{14,53} = h/(n+1)e^2$, i. e. the Hall
resistances are precisely quantized, whereas $R^L_{14,23} =
R^L_{14,65} = [h/(n+1)e^2]\cdot t_2/(n+1-t_2)$. Since $t_2$ has a
strong resonant dependence on $E_F$ (or $\nu$), it follows that $R_L$
fluctuates strongly. In particular, if $n=0$ (transition inside
spin-up LLL), $R_L$ can be arbitrarily large when $t_2\rightarrow 1$,
whereas in higher LLs the amplitude of fluctuations in $R_L$ is
$h/(n+1)e^2$ or less, as observed both experimentally and in our
simulations.

If $B$ changes sign, we have verified that the identity $\hat{g}(-B) =
[\hat{g}(B)]^T$ holds~\cite{Baranger}.  The reason is that
time-reversal symmetric tunneling is not affected by this sign change,
while the flow of the chiral currents is reversed. The model mirrors
itself with respect to the horizontal axis if $B$ changes sign, see
Fig.~\ref{fig1}.  The solutions of $\hat{I}_{14} = \hat{g}(-B) \cdot
\hat{v}$ are then related to the solutions of $\hat{I}_{14} =
\hat{g}(B) \cdot \hat{V}$ by $v_2 = V_6$, $v_3 = V_5$, $v_5 = V_3$ and
$v_6 = V_2$, provided that the same index exchanges $ 2
\leftrightarrow 6$, $ 3 \leftrightarrow 5$, are performed for all
$r_{ab}$.  Terms not invariant under this transformation are $r_{12}$,
$r_{16}$, $r_{43}$, $r_{45}$, $r_{23}$ and $r_{56}$. As already
discussed, the last two terms are vanishingly small. In the
experimental setup, the first four terms are also very small, due to
the long distance between source and drain, and their nearby
contacts~\cite{note1}.  The dominant terms $r_{26}$ and $r_{35}$ are
invariant under the index exchange. It follows then that
$R^L_{14,23}(B)= R^L_{14,65}(-B)$ and vice versa, i.~e. with good
accuracy, the fluctuation pattern of one $R_L$ mirrors that of the
other $R_L$ when $B \rightarrow -B$. This symmetry has indeed been
observed experimentally, with small violations at
low-$\nu$~\cite{private} due to perturbative corrections from the
non-invariant tunneling contributions $r_{12}-r_{16}$ and $r_{43}-
r_{45}$.

We now summarize our understanding of the various results of IQHE
measurements on mesoscopic samples. Similar to experiments, we find
that the transition in higher LLs is naturally divided in three
regimes. At low-$\nu$, the LL hosting $E_F$ is insulating and there
are no edge states connecting the left and right sides of the
sample. If tunneling between left and right sides is also small, we
find that the fluctuations of pairs of resistances are correlated with
excellent accuracy, $R_H +R_L = h/ne^2$. This condition is obeyed if
the typical size of the wave-function (localization length) is less
than the distance between contacts 2 and 3.  When the localization
length becomes comparable to this distance, an edge state is
established and the correlation between $R_L$ and $R_H$ is lost.  On
the high-$\nu$ side the edge states are established, but localized
states inside the sample can help electrons tunnel between opposite
edges, leading to back-scattering like the Jain-Kivelson model. In
this case, we showed that $R_L$ fluctuates while $R_H$ is
quantized. Tunneling between opposite edges is likely only if the
typical size of the wave-functions is slightly shorter than the
distance between opposite edges.  It is then apparent that the central
regime in Figs. \ref{fig3} (b) and (c) corresponds to the so-called
``critical region'', where the typical size of the electron
wave-function is larger than the sample size (distance between
contacts 2 and 3, at low-$\nu$, or between 2 and 6 at high-$\nu$).  In
these mesoscopic samples, the voltage probes act as markers on a
ruler, measuring the localization length of the wave-functions at the
Fermi energy.  To our knowledge, this is the first time when the
boundaries of the critical region are pinpointed experimentally.  This
opens up exciting possibilities for experimentally testing the
predictions of the localization theory.

To conclude, we used both first-principles simulations and a simple
model to explain the phenomenology of the mesoscopic IQHE, for
six-terminal geometry. We identified tunneling and chiral currents as
coexisting mechanisms for charge transport in mesoscopic samples, and
argued that the boundaries between the three distinct regimes mark the
critical region.

This research was supported by NSERC and NSF DMR-0213706. 
We thank Y. Chen, R. Fisch, R. N. Bhatt and
D. Shahar for helpful discussions.


\begin{thebibliography}{99}
\bibitem{note1} We use short samples ($\approx~4\mu$m$\times 2\mu$m)
  in order to save CPU time by decreasing the number of states in the
  LLL; experimental devices are much longer
  ($\approx~24\mu$m$\times2\mu$m).
\end{thebibliography}
\end{document}